\documentclass[prl,aps,twocolumn,groupedaddress,showpacs,floatfix]{revtex4}

\usepackage{amsmath,amssymb,multirow,epsfig,bm}

\newcommand{\beq}{\begin{equation}}
\newcommand{\eeq}{\end{equation}}
\newcommand{\bea}{\begin{eqnarray}}
\newcommand{\eea}{\end{eqnarray}}

\newcommand{\benn}{\begin{displaymath}}
\newcommand{\eenn}{\end{displaymath}}

\begin{document}

\title{Signature of the
electron-electron interaction in the magnetic field dependence of
nonlinear I-V characteristics in mesoscopic systems.}

\author{B.Spivak}

\affiliation{Physics Department, University of Washington,
Seattle, WA 98195, USA}

\author{A.Zyuzin}

\affiliation{A.F.Ioffe Institute, 194021 St.Petersburg, Russia}

\begin{abstract}
We show that the nonlinear I-V characteristics of mesoscopic  samples
 with metallic conductivity should contain parts which are linear in the
magnetic
field and
quadratic in the electric field. These contributions to the current
are entirely due to the
electron-electron
interaction and consequently they are  proportional to the
 electron-electron interaction
constant. We also note that both the amplitude and the sign
 of the current exhibit
random oscillations as a function of temperature.
\end{abstract}


\maketitle

According to the Onsager relation
the linear conductance $G({\bf H})$ of a conductor
 measured by the two-probe method must be an even function of the
 magnetic field ${\bf H}$ \cite{Landau}:
\begin{equation}
G({\bf H})=G(-{\bf H})
\end{equation}
 Eq.1 is a consequence of general principles: the time reversal
 symmetry and the positive sign of the  entropy production. Therefore it
 holds in all conductors.  It is possible however that that nonlinear I-V
characteristics of conductors contain parts odd in ${\bf H}$.
 In particular, one can have contributions to the total current
 through a sample which
 are  linear in $H$ and
quadratic in the voltage across the sample $V$.
\begin{equation}
I_{(nl)}=\alpha V^{2} H
\end{equation}
Since ${\bf H}$ is an axial vector and the current is a polar one,
 the coefficient $\alpha$ can be non-zero only in
 non-centrosymmetric media.
 In the case of bulk non-centrosymmetric crystals
 terms in
 $I-V$ characteristics that are linear in ${\bf H}$ have been
investigated both
 theoretically, the using Boltzmann kinetic equation, and experimentally
(See for example \cite{Photovoltaic}). In the case of chiral
carbon nanotubes a classical theory of this effect was discussed in
\cite{ivchenko}.

In this article we study this effect at small temperatures and in
 mesoscopic disordered samples where all possible symmetries are
broken. In this situation all electron transport effects are of
a quantum interference nature.
The theory of nonlinear
 characteristics of mesoscopic metallic samples was
 developed in the approximation of non-interacting electrons
\cite{LarkinKhmelnitskii,KhmelnitskiiFalko}.
It is important, however, that in this approximation
$\alpha=0$ and magnetic field dependence of the $I-V$
 characteristics is an even
function of
${\bf H}$.
Therefore the
 coefficient $\alpha$ in Eq.2 should be proportional to
electron-electron interaction constant $\beta$, which is
defined
 by the interacting part of the electron Hamiltonian
\begin{equation}
H_{(int)}=\frac{\beta}{\nu}
\int d {\bf r} \Psi({\bf r}) \Psi^{*}({\bf r}) \Psi({\bf r})
\Psi^{*} ({\bf r})
\end{equation}
Here $\nu$ is the electron density of states.
Thus, in principle, by measuring the current in Eq.2 one can measure the
electron-electron interaction constant $\beta$.

Let us consider a sample of two-dimensional geometry shown in
the insert of Fig.1
and assume that the magnetic field is perpendicular to the plane and that
the characteristic size of the sample $L\gg l$
is much larger then the electron elastic mean free path $l$.
 At low temperatures
the main contributions to both mesoscopic fluctuations
of the conductance $\delta G=G-\langle G \rangle$ and the nonlinear
 current Eq.2
are due to electron interference effects. As usual in such situations $\alpha$
is random sample specific quantities
 with zero average $\langle \alpha \rangle =0$.
To characterize $\alpha$ one has to calculate the variance
$\langle \alpha^{2} \rangle$. Here the brackets $\langle
\rangle$ denote averaging over realizations of a random white noise
scattering
 potential characterized by a correlation function
 $\langle u({\bf r}) u({\bf r}) \rangle =
\frac{\pi}{lm^{2}}\delta({\bf r-r'})$, where $m$ is the
electron mass.

A diagrammatic verification of Eq.1 in the situation when interference
 corrections to conductance are significant  is not entirely trivial
 and involves
 calculation of diagrams
shown in Fig.1a,b \cite{AltshulerKhmelnitski}. We will use a standard
diagram technique for averaging over random realizations of the scattering
potential \cite{Abricosov}.
 To verify that the correlation function of
 conductances $\langle \delta G({\bf H})
\delta G(0)\rangle$ is an even function of the magnetic field
one has to take into account contributions to the correlation function of
"Diffuson" and "Cooperon" propagators which are represented by the
ladder
diagrams in Fig.1 a,b.
(The "Diffuson" is the ladder part of the diagram in Fig.1a where the
direction
of arrows of the electron Green function is antiparallel, while the
"Cooperon" is the ladder in Fig.1b with parallel directions of the arrows.
They
 contain parts linear in the magnetic field, which are
equal in magnitude and of different signs. Thus these contributions
cancel.
To verify the fact that $\alpha=0$ in  the approximation of
non-interacting
 electrons
one has to calculate diagrams shown in Fig.1c. In this case the linear in ${\bf
H}$ the Diffuson and
the Cooperon contributions will cancel each other as well.
This fact is also
quite obvious in the framework of Landauer scheme of calculations
 of the conductance.

In linear in $\beta$ approximation the variance
\begin{equation}
\langle \alpha^{2} \rangle=
\beta^{2}\frac{e^{2}}{\nu^{2} \Gamma^{4}A^{2}}
(\frac{e^{2}}{h})^{2}
(\frac{L^{2}}{\Phi_{0}})^{2}
\end{equation}
is given by diagrams shown in Fig.1d.
Here $\Gamma=\hbar D/L^{2}$ and $D/L^{2}$ is the inverse lifetime
of
an electron in the sample, $A$ is the area of the
sample, and $D=v_{F}l/2$ is the electron diffusion coefficient.
Eq. 4 is valid at $eV\ll \Gamma$ and $\Phi\ll \Phi_{0}$, where
$\Phi=HA$ and $\Phi_{0}$ are the magnetic flux through the sample
area and the flux quanta respectively.
For simplicity we consider the short range e-e interaction described by
 Eq.3. In this case
the Hartree term is twice larger than the exchange term and we consider
only the
diagram shown in Fig.1d.

In the case of the Coulomb interaction between electrons and at high
 electron densities we have
\begin{equation}
\beta= e^{2} r_{D}\nu
\end{equation}
where $r_{D}$ is the screening radius ( $e^{2}r_{D}/D\sim 1/k_{F}l$).

The existence of the effect described by Eq.4 is connected to the
 fact that in the presence
of a voltage $V$ across the sample there is a part of the
local density
\begin{equation}
\delta n({\bf r})\sim  VH
\end{equation}
which is proportional to $H$ and $V$ \cite{SpivakZyuzin}.
Existence of fluctuations of the density of the form
Eq.6 is a consequence of the fact
that linear in ${\bf H}$ parts of the "Diffusion" and "Cooperon"
digrams shown in Fig.1d do not cancel.
 For our estimate it is enough to consider only "Diffuson"
contribution.

We would like to stress that the effect described by Eq.4 is quite
different from conventional effects in bulk crystals which
 can be described by the
Boltzmann kinetic
equation \cite{Photovoltaic,ivchenko}.  The leller effects are determined
by
relaxation processes in materials with complicated band structures and for
this reason they are proportional to the the relaxation rates
(or proportional to $\beta^{2}$), while Eq.2 is proportional to $\beta$.

The qualitative explanation of Eq.4 is the following.
The mesoscopic fluctuations of the current density
inside the sample are due to random interference of electron waves
traveling along different diffusive paths. Though, the total current
through the sample should be an even function of ${\bf H}$
the local current densities contain a part proportional to ${\bf H}$.
For example, a part of the current density  proportional to ${\bf H}$
can be
characterized as a "Hall current density". To avoid confusion we
would
like to mention that
this "Hall component" is connected with the electric field in highly
non-local way and has a random direction.
By the same token in random system there is a component of electron
density described by Eq.6.
We note also that density fluctuations Eq.6 are different from Friedel
oscillations in disordered samples, which are even function of
${\bf H}$.
 In the Hartree
approximation there is an additional scattering potential
\begin{equation}
\delta u_{e}({\bf r})=\frac{\beta}{2\nu} \delta n({\bf r})
\end{equation}
associated with the fluctuations of the electron density.
Thus we can write an expression for a total current in the form
\begin{equation}
I=G [V,T, H, \{\delta u_{e}({\bf r}, H)\}] V
\end{equation}
where the nonlinear conductance $G= G_{D} + \delta G(V,T, H, \{\delta
u_{e}({\bf r}, H)\})$, generally
speaking, depends on the
realization of $u_{e}({\bf r})$ via the corresponding dependence of the
mesocopic
part $\delta G$. Here $G_{D}=e^{2} D \nu$ is the Drude conductance.

The sensitivity of the sample conductance to a change in
 the scattering potential
 $\delta u_{e}({\bf r})$ has been considered in \cite{LeeStone,AltshulerSpivak}.
 Generally speaking, the mesoscopic part of the conductance $\delta G$
depends on all
spatial harmonics of $\delta u_{e}({\bf r})$. However,
the main contribution to the change of the conductance comes from
zero harmonics of the potential
\begin{equation}
\delta \bar{u}_{e}(V, {\bf H})=\frac{\beta}{\nu A} \int (\delta n({\bf
r},V,H) -
\delta n({\bf r}, V, 0)) d {\bf r}
\end{equation}
This can be verified by making calculations similar to
those in \cite{SpivakZyuzinNL}. This is also related to the
long range character of the correlation function of the part of the
electron densities, which are proportional to $V$ \cite{SpivakZyuzin}. For
example, in the 2D case the correlation
function described by digrams shown in Fig.1c has the form
\begin{eqnarray}
\langle \delta u_{e}({\bf r}) \delta u_{e}({\bf r'})
\rangle=\frac{\beta^{2}}{\nu^{2}}\langle \delta n({\bf r}) \delta
n({\bf r}') \rangle = \nonumber \\
-\frac{\beta^{2}}{2\pi\nu^{2}}
(\frac{V}\pi^{2} h D)^{2} \ln \frac{|{\bf r-r'}|}{L}
\end{eqnarray}

Expanding Eq.8
($\delta I=(d \delta G/ d \bar{u}_{e}) \bar{u}_{e} V $) in terms of
$\bar{u}_{e}$ and
taking into account that in the main approximation $\bar{u}$ and
$\delta G$ are uncorrelated we get
\begin{equation}
\langle I^{2}_{(nl)} \rangle =\langle (d G/ d\bar {u}_{e})^{2}
 \rangle \langle \bar{u}_{e}^{2}\rangle
V^{2}
\end{equation}
According to \cite{AltshulerSpivak}
$\langle(dG/du)^{2}) \rangle=(e^{2}/h)^{2}/\Gamma^{2}$.
Calculating the correlation function
\begin{equation}
\langle \delta
(\bar{u}_{e}(H))^{2}\rangle=
\frac{\beta^{2}}{\nu^{2}}\frac{1}{A^{2}}\frac{\Phi^{2}}{\Phi_{0}}
\frac{|eV|^{2}}{\Gamma}
\end{equation}
we arrive at Eq.4.

At this point we would like to mention that on a qualitative level
the effect considered above can be also described in the framework
of Landauer-Buttiker scheme. To do so one has to
 combine results of
\cite{Butiker} and \cite{Butiker1}.

Eq.4 is valid at small temperatures $\Gamma \gg T$. At finite temperature
the quantity $\langle \alpha^{2}(T) \rangle$ decreases with $T$.
At $T\gg \Gamma$
\begin{equation}
\langle \alpha^{2}(T) \rangle \sim \langle \alpha^{2}(0) \rangle
\frac{\Gamma^{2}}{T^{2}}
\end{equation}

We stress that the temperature dependence of $\alpha(T)$
 is non-monotonic: $\alpha(T)$ exhibits random oscillations in
magnitude and  sign, superimposed on the average decay.
One can see this by calculating the quantity
\begin{equation}
\langle \alpha (T)\alpha (0)\rangle \sim \langle \alpha^{2}(0) \rangle \frac{\Gamma^{2}}{T^{2}}
\end{equation}
Note that Eqs.13,14 have the same temperature dependence, which is
impossible without oscillations of the sign
of $\alpha(T)$ \cite{conductance}.

In the case of high magnetic field $\Phi > \Phi_{0}$, (but  still
$eV\ll\Gamma$), the part of the current which is  the asymmetric in $H$
and
quadratic in $V$ exhibits random oscillations as a function of
$\Phi$.
These oscillations, typical for mesoscopic systems have characteristic
period
$\Phi_{0}$ and the amplitude
\begin{equation}
\langle (I({\bf H})-I({\bf -H}))^{2}
\rangle =\beta^{2}(\frac{eV}\Gamma)^{2}\frac{1}{(\nu \Gamma A)^{2}}
(\frac{e^{2} V}{h})^{2}
\end{equation}

Finally, we mention that there are no mechanisms contributing
to Eq.2 other than  the mechanism considered above.
For example, at finite $V$ there is a new channel of
electron
transmission through the sample when an incident electron
is transmitted into two electrons and a hole.
The probability of such a process has a component which is linear in $H$.
It's magnitude can be
estimated in a way similar to the estimating the electron-
 electron scattering rate of quasiparticles in a
 uniform Fermi liquid. As a result, it is proportional to $V^{2}$.
Thus the magnitude of the asymmetric-in-$H$ part of
 the current associated with such process is proportional to $V^{3}H$.

Recently discussed above effect was observed experimentaly
in $GaAs$ quantum dots \cite{Markus}.

During the preparation of the manuscript we  became aware of similar
unpublished results by D. Sanchez and M. B\"uttiker \cite{Sanchez}.

This work was supported in part by the National Science
Foundation under Contracts No. DMR-0228104.
We thank M. Buttiker, C. M. Marcus and D. M. Zumbuhl, for
useful discussions.

\newpage

\begin{figure}
  \centerline{\epsfxsize=10cm \epsfbox{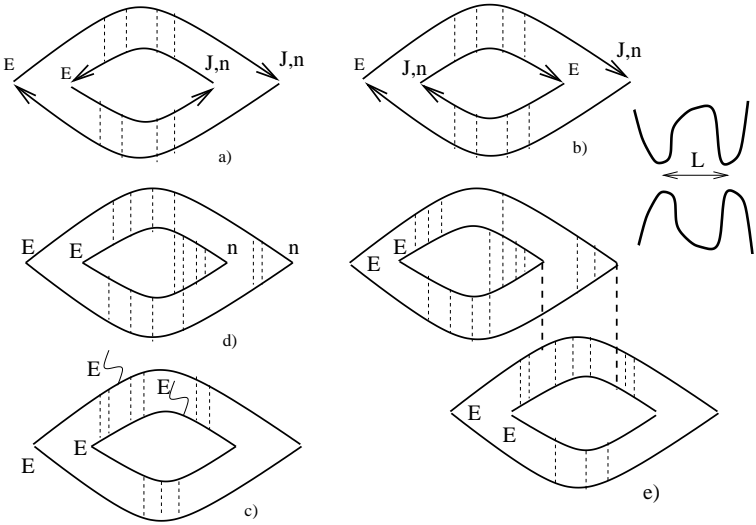}}
  \caption{ Solid lines correspond to
electron
Green's functions, thin dashed lines correspond to the correlation
function
of the
random scattering potential $\langle u({\bf r}) u({\bf r'})\rangle$ ,
thick dashed lines correspond to the
electron-electron interaction $\beta/\nu$. Symbols $E, J,n$ correspond to
electric field, current density and electron density respectively.
Diagrams a) and b) describe
the correlation function $\langle G({\bf H}) G(0) \rangle$. Parallel and
antiparallel directions of arrows in the ladder parts of these diagrams
 correspond to the Cooperon and Diffusons respectively.
The diagram c) describes the mesoscopic fluctuations of the part of the
electron density which
is proportional to the electric field squared in the approximation of
noninteracting electrons. The diagram d) describes Eq.10. The diagram
e) describes Eqs.4, 13, 14.
The insert shows a schematic picture of the sample. } \
  \label{fig:fig1}
\end{figure}


\begin{thebibliography}{99}

\bibitem{Landau} L. D. Landau E. M. Lifshitz, Statistical physics,
Butterworth, Heinemann, (1980).
\bibitem{Photovoltaic} B.I.~Sturman, V.M.~Fridkin, The Photovoltaic and
Photorefractive effects in Non-centrosymmetric Materials, Gordon and
Breach
Science Publishers, (1992); G. L. J. A. Rikken, J. Fölling, and P. Wyder
Phys.Rev.Lett. {\bf 87}, 236602, (2002).
\bibitem{ivchenko} E. Ivchenko, B. Spivak, Phys. Rev. B {\bf 66}, 155404,
(2002).
\bibitem{LarkinKhmelnitskii} A. Larkin, D. Khmelnitskii, JETP,
{\bf 91}, 1815, (1986);
Physics Scripta, T14, 4, (1986).
\bibitem{Abricosov} A.A. Abrikosov, L.P. Gorkov, I. Dzyaloshinski,
"Methods of Quantum Field Theory in Statistical Physics"Dover Publ. Inc. NY 1975.
\bibitem{KhmelnitskiiFalko} V.Falko, D.Khmelnitskii, Sov.Phys.JETP, {\bf
68}, 186, (1989).
\bibitem{AltshulerKhmelnitski} B. L. Altshuler, D. E. Khmelnitskii,
JETP Lett, {\bf 42}, 359, (1985).
\bibitem{SpivakZyuzin} A. Zyuzin, B. Spivak,
Sov.Phys.JETP.{\bf 66}, 560-565 ,(1987); B. Spivak, A. Zyuzin,
"Mesoscopic Fluctuations of Current Density
in Disordered Conductors" In "Mesoscopic Phenomena in Solids" Ed.
B. Altshuler, P. Lee, R. Webb, Modern Problems in Condensed Matter
Sciences vol.30, (1991).
\bibitem{SpivakZyuzinNL} B. Spivak, A. Zyuzin, J. Opt. Soc. Am. {\bf B21},
177, (2004).
\bibitem{LeeStone} P.A. Lee, A.D. Stone,
 Phys.Rev.Lett.{\bf 55}, 1622-1625, (1985).
\bibitem{AltshulerSpivak} B. Altshuler, B. Spivak, JETP Lett.{\bf 42},
447-449, (1986).
\bibitem{Butiker} M. B\"{u}ttiker,
J. Phys. Condensed Matter {\bf 5}, 9361 - 9378, (1993).
\bibitem{Butiker1} M. B\"uttiker, Europhys. Lett. {\bf 35}, 523 (1996).
\bibitem{conductance} We note that similar
phenomenon exist in the case of mesoscopic oscillations
of linear conductance. Namely the quantity $\Delta G=(G(H,T)-G(0,T))$
exhibits random oscillations both in magnitude and in sign as a function
of
$T$. One can see this by calculating the variances
 $\langle (\Delta G(T))^{2}\rangle \sim \langle \Delta G(T) \Delta
G(0) \rangle \sim 1/T$. Both quantities have the same $T$-dependence,
 which is impossible if sign $\Delta G(T)$ does not oscillate as a
function
of $T$. As far as we know this fact
has not been discussed in the literature.
\bibitem{Sanchez} D. Sanchez, M. B\"uttiker, Unpublished.
\bibitem{Markus} D. M. Zumbuhl, C. M. Marcus, M. P. Hanson, A. C. Gossard;
In  preparation.


\end{thebibliography}
\end{document}